\documentclass[twocolumn,showpacs,10pt,superscriptaddress,floatfix,longbibliography]{revtex4-2}
\usepackage{graphicx}
\usepackage{amssymb}
\usepackage{bm}
\usepackage{amsthm}
\usepackage{amsmath}
\usepackage{xcolor}
\usepackage{hyperref}        
\hypersetup{
	colorlinks=true,
	citecolor=blue,
	urlcolor=blue}

\begin{document}
\title{Inverse ac Josephson effect in Josephson diode}
\author{Huarong Zhong}
\affiliation{  School of Physics, Sun Yat-sen University, Guangzhou 510275, China}
\author{Zhi Wang}
\email{wangzh356@mail.sysu.edu.cn}
\affiliation{  School of Physics, Sun Yat-sen University, Guangzhou 510275, China}
\affiliation{ Guangdong Provincial Key Laboratory of Magnetoelectric Physics and Devices, Sun Yat-sen University, Guangzhou 510275, China}

\begin{abstract}
We study the Josephson junction where nonreciprocal critical current was induced by the interplay of the $4\pi$-periodic and $2\pi$-periodic current-phase relation of the junction. We take the model of a topological junction which serves as a Josephson diode with nonreciprocal critical currents. For this Josephson diode, we demonstrate an inverse ac Josephson effect where an effective dc voltage is induced by a pure ac driving current. We show that this inverse ac Josephson effect originates from the voltage rectification by the nonreciprocal critical current of the system. We explore the dependence of the induced dc voltage on the amplitude and frequency of the ac driving current and reveal the optimized condition for the inverse ac Josephson effect.
\end{abstract}
\maketitle

\section{Introduction}
Josephson effect is a quantum tunneling phenomenon where Cooper pairs tunneling coherently between two superconductors\cite{Josephson1974}. The quantum coherence of the tunneling process brings in two distinctive features: the dc Josephson effect where dc current can flow without voltage, and the ac Josephson effect where a dc biased voltage drives an ac current. The Josephson effects have been extensively investigated for understanding of quantum phenomenon in macroscopic scale\cite{Kwon2004EuropeanPhysical,Lei2018PhysRevLett.121.227701,Schrade2018PhysRevLett.120.267002,Razmadze2020PhysRevLett.125.116803,Misaki2021PhysRevB.103.245302,Kopasov2021PhysRevB.103.144520,Baumgartner2022Nature,Margarita2022Science,Pillet2023PhysRevResearch,Ciaccia2023PhysRevResearch,
Sun2023PhysRevB,Wei2023PhysRevB,Liu2024PhysRevB,Osin2024PhysRevB,Virtanen2024PhysRevLett,Cayao2024PhysRevB,
Schrade2024PhysRevApplied,Wang2024PhysRevB.109.075412,Seoane2024PhysRevResearch,Volkov2024PhysRevB}. Recently, it has received reviving interest due to the progress in quantum computation based on superconducting quantum bit and quantum gates\cite{Razmadze2020PhysRevLett.125.116803,Wang2022PhysRevLett.129.257001,James2022NewJournal,Zhang2022PhysRevX.12.041013,Souto2022PhysRevLett.129.267702,Cheng2023PhysRevB.107.184511,Debnath2024PhysRevB}. In particular, the superconducting quantum devices built by the Josephson junction is very attractive because of its advantage in the large-scale circuit construction and fast quantum man,ipulation\cite{Kamata2018PhysRevB.98.041302}.
The diode is an important ingredient in electronic circuits. A superconducting Josephson diode is potentially useful for quantum devices\cite{Osin2024PhysRevB,Vakili2024PhysRevB.110.104518}. By definition, a Josephson diode should have a nonreciprocal supercurrent\cite{Souto2022PhysRevLett.129.267702,Noah2022,Askerzade2022PhysicaC,Ortega2023PhysRevB,Awoga2024PhysRevResearch,Awoga2024PhysRevB,Debnath2024PhysRevB,Kumari2024PhysRevLett}. This requires the breaking of the time-reversal symmetry, which reverses the direction of the supercurrent\cite{Oostinga2013PhysRevX.3.021007,Iliifmmode2022PhysRevLett.128.177001}. In addition, the current phase relation of the junction should have multiple harmonics, because a proper phase shift of a single harmonic function is always able to reverse the sign of the supercurrent, as stated by the Euler identity. The interplay of the time-reversal symmetry breaking and the multiple harmonic current-phase relation thereby sets up one simple picture for understanding the Josephson diodes\cite{Hu2007PhysRevLett.99.067004,Recher2010PhysRevLett,Chen2018PhysRevB.98.075430,Chen2018PhysRevB.98.165439,Turini2022Nano,Askerzade2022PhysicaC,wang2022symmetryconstraint,Sivakumar2022Nature,Pal2022NaturePhysics,
Ortega2023PhysRevB,Matsuo2023NaturePhysics,Hu2023PhysRevLett.130.266003,Trahms2023Nature,Greco2023dc-SQUID,Gupta2023NatureCommunications,Costa2023PhysRevB,Steiner2023PhysRevLett,Dutta2024NatureMaterials,
Coraiola2024,Reinhardt2024NatureCommunications}. While the higher harmonics are well suited for generating Josephson diodes, one unique choice is the half harmonic current phase relation of the topological junctions\cite{Liu2019PhysRevLett.122.086804}. There have been several proposals along this direction, including the topological superconducting interference devices and topological long junctions\cite{Scharf2021PhysRevResearch.3.033062,Jian2021PhysRevB.103.134514}.

Topological junctions exhibit unique transport behaviors due to the inherent Majorana qubit\cite{Wang2022PhysRevLett.129.257001,Kitaev2001,Awoga2024PhysRevB,Kumari2024PhysRevLett,Awoga2024PhysRevResearch}, such as intrinsic hysteresis and suppression of the first Shapiro step\cite{Li2018PhysRevB.97.045423}. These behaviors are well described by the quantum resistively and capacitively shunted junction model\cite{2004tinkham}. In this work, we propose a topological diode with the simplest structure, which combines the $4\pi$ periodic Majorana junction and the conventional $\varphi$ junction\cite{Kwon2004EuropeanPhysical,Kitaev2001,Nakata2024PhysRevResearch.6.033207}. We show how the interplay of the symmetry breaking and the multi-harmonic current phase relation induces a nonreciprocal critical current in the junction\cite{Ryohei2017ScienceAdvances,Hoshino2018PhysRevB.98.054510}. We focus on the transport of the junction with a current injection and reveals an inverse ac Josephson effect, that is, a pure ac injected current can drive a dc voltage. We show that the voltage rectification by the nonreciprocal current is the central ingredients from this inverse ac Josephson effect. We explore the parameter space of the junction to identify the optimal condition for the inverse Josephson effect, and demonstrate typical junction dynamics for both overdamped and underdamped junction.

\begin{figure}
	\centering
	\includegraphics[width=0.9\columnwidth]{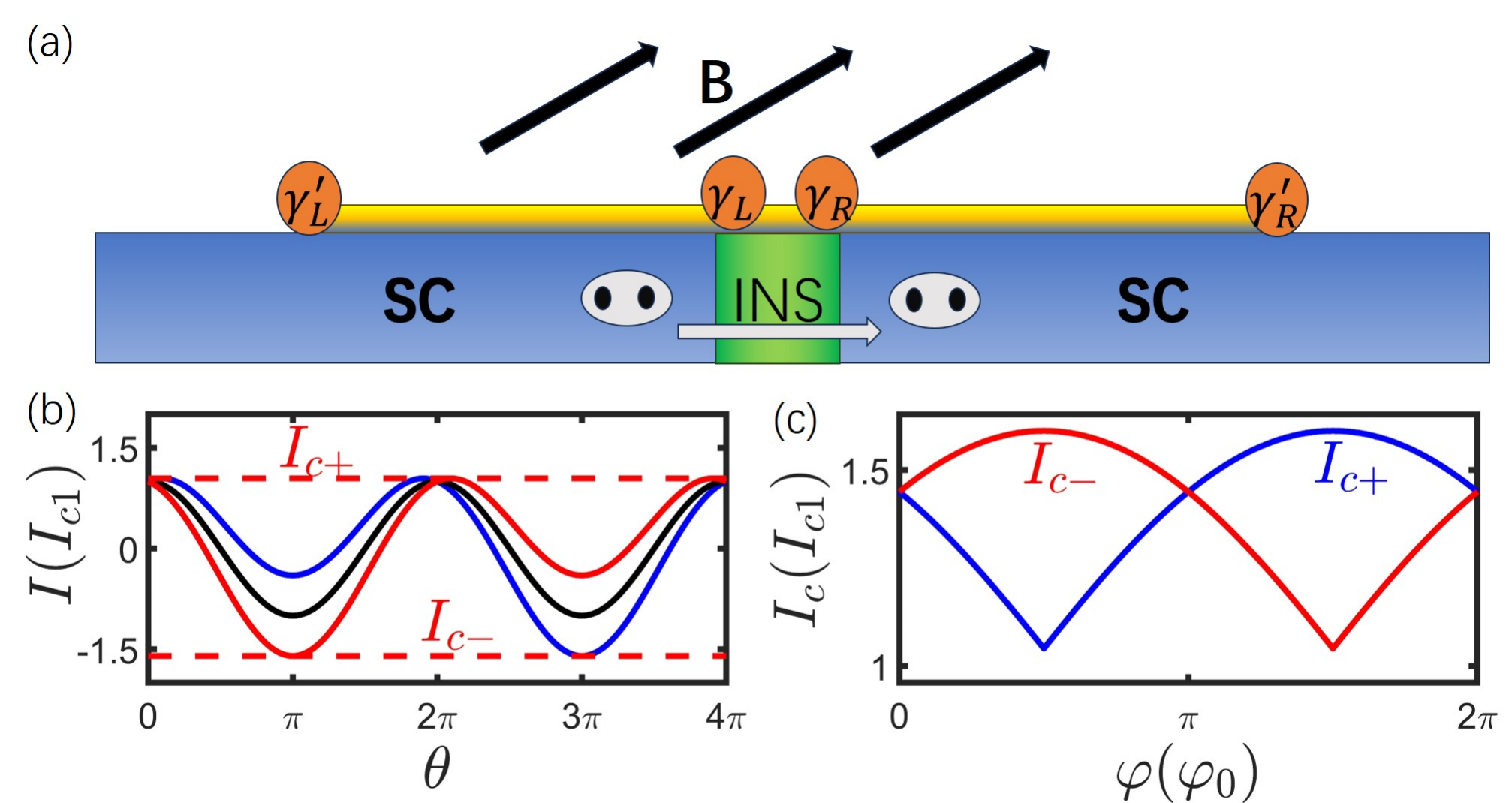}
	\caption{$(a)$ Schematics of a Topological Josephson diode, which consists of a one dimensional spin-orbit coupling system in proximity to two conventional s-wave superconductors. Majorana zero modes appear in presence of Zeeman energy and the two Majorana zero modes carry a $4\pi$-periodic Josephson current which has a finite phase shift $\varphi$ when the direction of the Zeeman field is not parallel to the one dimensional system. The conventional $4\pi$-periodic Josephson current can be transported between two s-wave superconductors. $(b)$ Current phase relation of the system for different phase shift. The nonreciprocal current-phase relation comes from the interplay of the $4\pi$-periodic and the $2\pi$ period current-phase relation. $\varphi/\varphi_{0}={\pi}/{2}$, where $\varphi_{0} = {\hbar}/{2e}$ is the flux quantum. $(c)$ The corresponding critical current as a function of the phase shift. Here we take the absolute value for the two critical currents in two directions for comparison. The critical current is directional asymmetric except for phase shifts that a integer number of $\pi$. }
	\label{fig:model}
\end{figure}

This work is organized as follows. In section II, we show the model of the proposed system and analyze the origin of the nonreciprocal current phase relation. In section III, we calculate the time evolution of the Josephson phase and the average dc voltage in the presence of a proper ac current, and demonstrate the inverse ac Josephson effect for typical junction parameters. In section IV, we explore the parameter space of the junction to identify the optimized parameters for the inverse ac Josephson effect. In section V, we give discussions and a summary.

\section{Model of a Majorana Josephson Diode}
The Majorana Josephson diode consists of a conventional Josephson junction in parallel to a topological junction\cite{Oreg2010PhysRevLett.105.177002,Lutchyn2010PhysRevLett.105.077001}. The minimal structure is illustrated in Fig. \ref{fig:model}a, where a one-dimensional topological superconductor, such as spin-orbit coupling nanowire with Zeeman energy, is placed upon two conventional s-wave superconductors\cite{Setiawan2017PhysRevB.95.174515}.The Cooper tunneling through the insulating barrier between the s-wave superconductors produce the usual $2\pi$ periodic Josephson current with a current phase relation of $I = I_{c_1} \sin \theta$ with a superconducting phase difference $\theta$ across the junction\cite{2004tinkham}. In contrast, the supercurrent through the nanowire is highly unconventional. To describe wire, we can introduce a tight-binding Hamiltonian for the superconductors on the two sides of the junction as\cite{Huang2017PhysicaC},
\begin{align}
         &H_\alpha(\phi_\alpha)=-t \sum_{\langle i,j\rangle,\alpha,\sigma} c^\dagger_{i,\alpha,\sigma} c_{j,\alpha,\sigma} - \mu \sum_{i,\alpha,\sigma} c^\dagger_{i,\alpha,\sigma} c_{i,\alpha,\sigma}\nonumber\\
&+\sum_{i,\alpha}{\Delta e^{i\phi_{\alpha}}c^{\dagger}_{i,\alpha,\uparrow}c^{\dagger}_{i,\alpha,\downarrow}}+\eta \sum_{i,\alpha,\sigma,{\sigma}^{\prime}}^{n}[c^{\dagger}_{i+1,\alpha,\sigma}(i{\sigma}_y)_{\sigma,{\sigma}^{\prime}}c_{i,\alpha,{\sigma}^{\prime}}]\nonumber\\
&+\sum_{i,\alpha,\sigma,{\sigma}^{\prime}}^{n}c^{\dagger}_{i,\alpha,\sigma}({\boldsymbol V}\cdot {\boldsymbol \sigma})_{\sigma,{\sigma}^{\prime}}c_{i,\alpha,{\sigma}^{\prime}},
\label{eq:H_w}
 \end{align}
where $\alpha$ labels the left and the right superconducting systems and $
\phi_{\alpha}$ represents the superconducting phase. For simplicity we take identical parameters for the two segments with $t$ the nearest-neighbor hopping, $\mu$ the chemical 
potential, $\eta$ the spin-orbit coupling, $\Delta$ the amplitude of the superconducting gap from the proximity effect, and $\boldsymbol V\cdot {\boldsymbol \sigma}$ the Zeeman energy from the Zeeman field $\boldsymbol V$. These two segments are connected through the electron tunneling Hamiltonian,
\begin{equation}
H_T=T\sum_{\sigma}(c^{\dagger}_{L,0,\sigma}c_{R,0,\sigma}+c^{\dagger}_{R,0,\sigma}c_{L,0,\sigma}),
\label{H_T}
\end{equation}
where $T$ is the tunneling energy. This Hamiltonian has been investigated and shown that the nanowire becomes a topological Josephson junction with four Majorana zero modes as shown in Fig. \ref{fig:model}a.
 There are two Majorana zero modes $\gamma_{L}$ and $\gamma_{R}$ situated near the tunneling barrier of the junction. These two Majorana zero modes construct a Majorana qubit with the eigenstates of the qubit defined by the opposite fermionic parity state of $i\gamma_{L}\gamma_{R}\left|\uparrow\right\rangle =\left|\uparrow\right\rangle$ and $i\gamma_{L}\gamma_{R}{\ensuremath{\left|\downarrow\right\rangle }}=-{\ensuremath{\left|\downarrow\right\rangle }}$\cite{2004tinkham}. In addition, there should be at least two other Majorana zero modes $\gamma_{L}^{\prime}$
and $\gamma_{R}^{\prime}$ at the end of the one-dimensional topological superconductor. Their coupling with the junction Majorana zero modes would flip the parity state of the qubit. These Majorana zero modes carries Josephson current with $4\pi$-periodic current-phase relation in the presence of proper horizontal Zeeman field\cite{Kitaev2001,Lutchyn2010PhysRevLett.105.077001,Alicea2011NaturePhysics}. 
Aside from the $4\pi$-periodic Josephson current, the Majorana zero modes in the nanowire junction also construct a Majorana qubit whose dynamics may significantly influence the transport of the junction\cite{Lutchyn2018NatureReviewsMaterials,Feng2018PhysRevB.98.134515,Stern2019PhysRevLett.122.107701,Li2019PhysRevB.99.100504,Feng2020PhysRevB.101.180504}. 
Taking the Josephson phase and the Majorana qubit into account together, we can arrive at the minimal Josephson Hamiltonian of the junction as\cite{Wang2022PhysRevLett.129.257001},
\begin{equation}
H_{M}=-E_{J}{\cos}({\theta+\varphi}) - E_{M}\sigma_{z}{\cos}(\frac{\theta}{2} )+E_{M}^{\prime}\sigma_{x}\label{eq:H_M},
\end{equation}
where $E_{J} = \hbar I_{c1} / 2e$ is the Josephson energy for the Cooper pairing tunneling, $E_{M} = T/4 $ is the Josephson energy due to the direct coupling between two Majorana zero modes around the junction\cite{Huang2017PLA}, $E_{M}^{\prime}$ is the coupling energy between the junction Majorana zero modes and the distant Majorana zero modes at the end of the wire and $\varphi$ is the relative phase shifted induced by the direction of the Zeeman field $\boldsymbol V$, which can be determined by numerically solving the tight-binding Hamiltonian\cite{Huang2017PhysicaC}, $\sigma_{z}$ and $\sigma_{x}$ represent pseudo-spin operators defined by the Majorana zero modes.

This Josephson Hamiltonian provides a nonreciprocal Josephson current which can be easily identified if we ignore the dynamics of the Majorana qubit by setting the $\sigma_z=1$. Then the Josephson current of the total junction is given as $I=I_{c1}\sin(\theta+\varphi)+I_{c2}\sin({\theta}/2)$, where $I_{c1}={2eE_{J}}/{\hbar}$ is the supercurrent from the Cooper pair tunneling and $I_{c2}={eE_{M}}/{\hbar}$ is the supercurrent from the half-pair tunneling through the Majorana
qubit. We plot this current phase relation for three typical phase shifts in Fig. \ref{fig:model}b, and find explicit directional asymmetry. To reveal the nonreciprocal Josephson current more explicitly, we calculate the critical current for the two directions as a function of the phase shift, as shown in Fig. \ref{fig:model}c. We find that the critical current is directional asymmetric, except for $\varphi = n\pi$. The critical current must be symmetric for $\varphi = n\pi$, because the Josephson current flips sign by reversal of the Josephson phase.

\begin{figure}
	\centering
	\includegraphics[width=0.9\columnwidth]{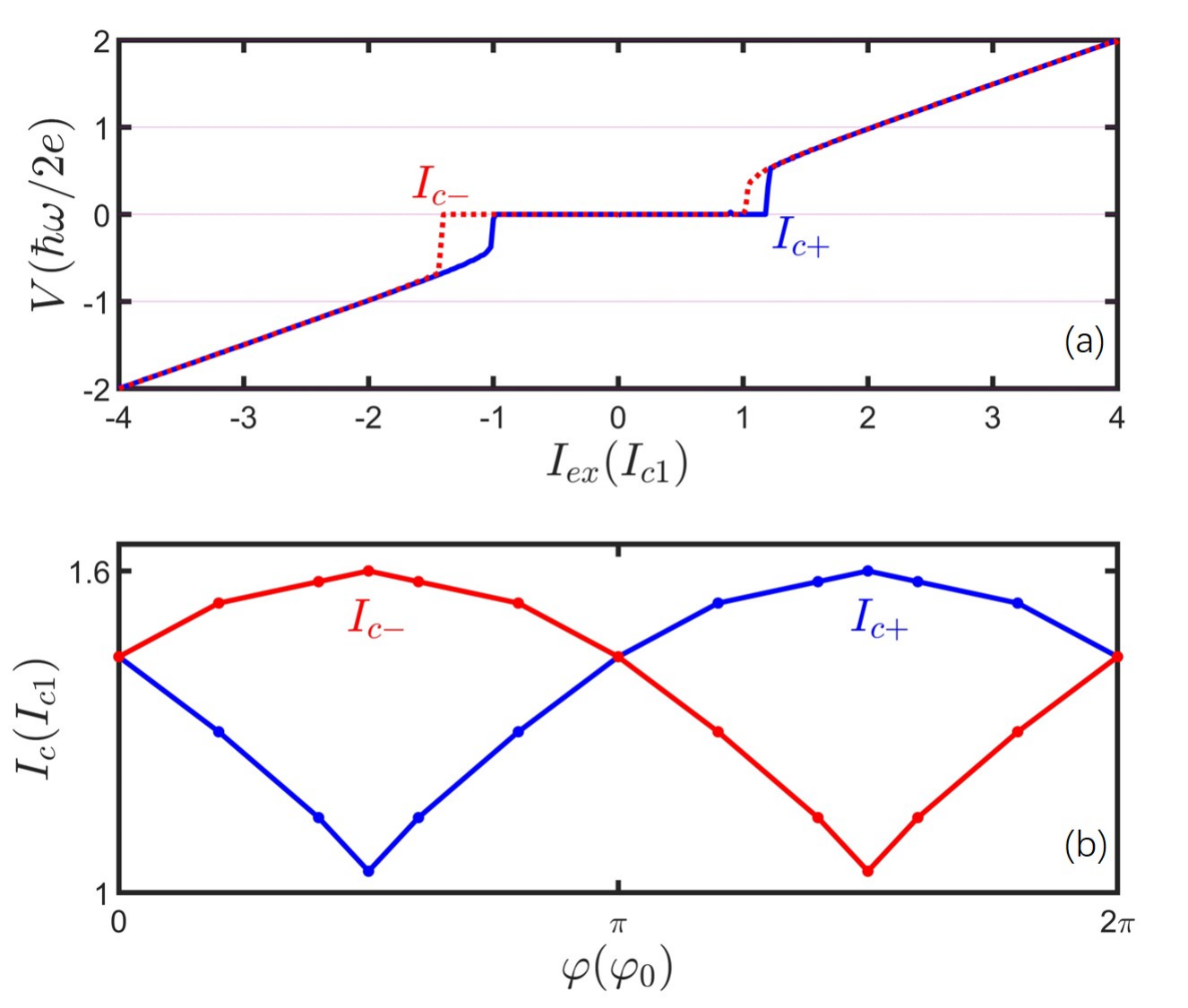}
	\caption{$(a)$ The I-V curve of the junction which exhibits hysteresis behavior and nonreciprocal critical current. $(b)$ The critical current of the two directions as a function of the phase shift between the $2\pi$-periodic and $4\pi$-periodic Josephson current. Here we take the absolute value for the two critical currents in two directions for comparison.}
	\label{fig:2}
\end{figure}

The Josephson Hamiltonian contains two degrees of freedom: the Josephson phase $\theta$ and the pseudo-spin state of the Majorana qubit. A proper dynamical equation for the transport of the junction should contain the time evolution of both degrees of freedom. This has been well described by the quantum resistively and capacitively shunted junction model. In this model, the dynamics of the Josephson phase is taken to be described by the classical equation of motion for a particle moving under a washboard potential, and the quantum states of the Majorana qubit enters the equation as a component for the effective potential. Meanwhile the dynamics of the Majorana qubit is described by the effective Landau-Lifshiz-Gilbert equation where the Josephson phase enters as a pseudo magnetic field. This model has been derived from the minimal Hamiltonian based on a path integral approach\cite{Wang2022PhysRevLett.129.257001}, and the resulted equation of motion is given as,
\begin{align}
    I_{ex} & =C\ddot{\theta}+\frac{\dot{\theta}}{R}+I_{c1}\sin(\theta+\varphi)+I_{c2}{s_{z}}\sin(\frac{\theta}{2})+\mathbf{B}_{f}\cdotp\dot{\mathbf{s}}\label{eq:action_current},\\
\dot{\mathbf{s}} & =\mathbf{h\times s}+\dot{\theta}{\mathbf{B}_{f}\times\mathbf{s}}+\left({\tilde{\mathbf{N}}}\cdot\dot{\mathbf{s}}\right)\times\mathbf{s}\label{eq:action_spin},
\end{align}
where $I_{ex}$ represents the adjustable current applied in the experiment, $\theta$ is the Josephson phase, $I_{c1}={2eE_{J}}/{\hbar}$
is the supercurrent from the Cooper pair tunneling and $I_{c2}={eE_{M}}/{\hbar}$
is the supercurrent from the half-pair tunneling through the Majorana
qubit. $C$ and $R$ are effective capacitance and resistance, $\mathbf{s}$
is the pseudo-spin and $x,y,z$ represents the corresponding components
in three directions respectively. $\mathbf{h}=({{E}_{M}^{\prime}},\textrm{0},{-{E}_{M}\cos({\theta}/2)})$
is the Zeeman field, ${\mathbf{B}_{f}}$ is the environment-mediated coupling field between the Josephson phase and the qubit, and ${\tilde{\mathbf{N}}}$ represents the environment-induced dissipation
to the qubit.

With these equations of motion, one can verify the Josephson diode behaviors with dc injected current\cite{Wang2012arXiv}. We can calculate the I-V characteristics of the junction with $I_{ex}$ which adiabatically increases from zero to a finite current and then decreases to zero, as shown in Fig. \ref{fig:2}a.
We notice that there is an intrinsic hysteresis in the I-V curve even though we are considering an overdamped junction. This is due to the influence of time evolution of the pseudo-spin $s_{z}$. As a result, we can define a "critical current" which is maximum current with zero voltage in the current increasing stage (blue line). We can also define a "retrapping current" which is the maximum current with zero voltage in the current decreasing stage (red line). From the I-V curve, we can extract the critical current and the retrapping current of the junction for the two directions in the I-V curve. We plot the critical currents of two directions as a function of the phase shift in Fig. \ref{fig:2}b, and find that the results resemble the one indicated in Fig. \ref{fig:model}b. This is understandable because the critical current of the I-V curve is determined by the maximum Josephson current with variations on both the Josephson phase and the Majorana qubit. 

\begin{figure}
	\centering
	\includegraphics[width=1\columnwidth]{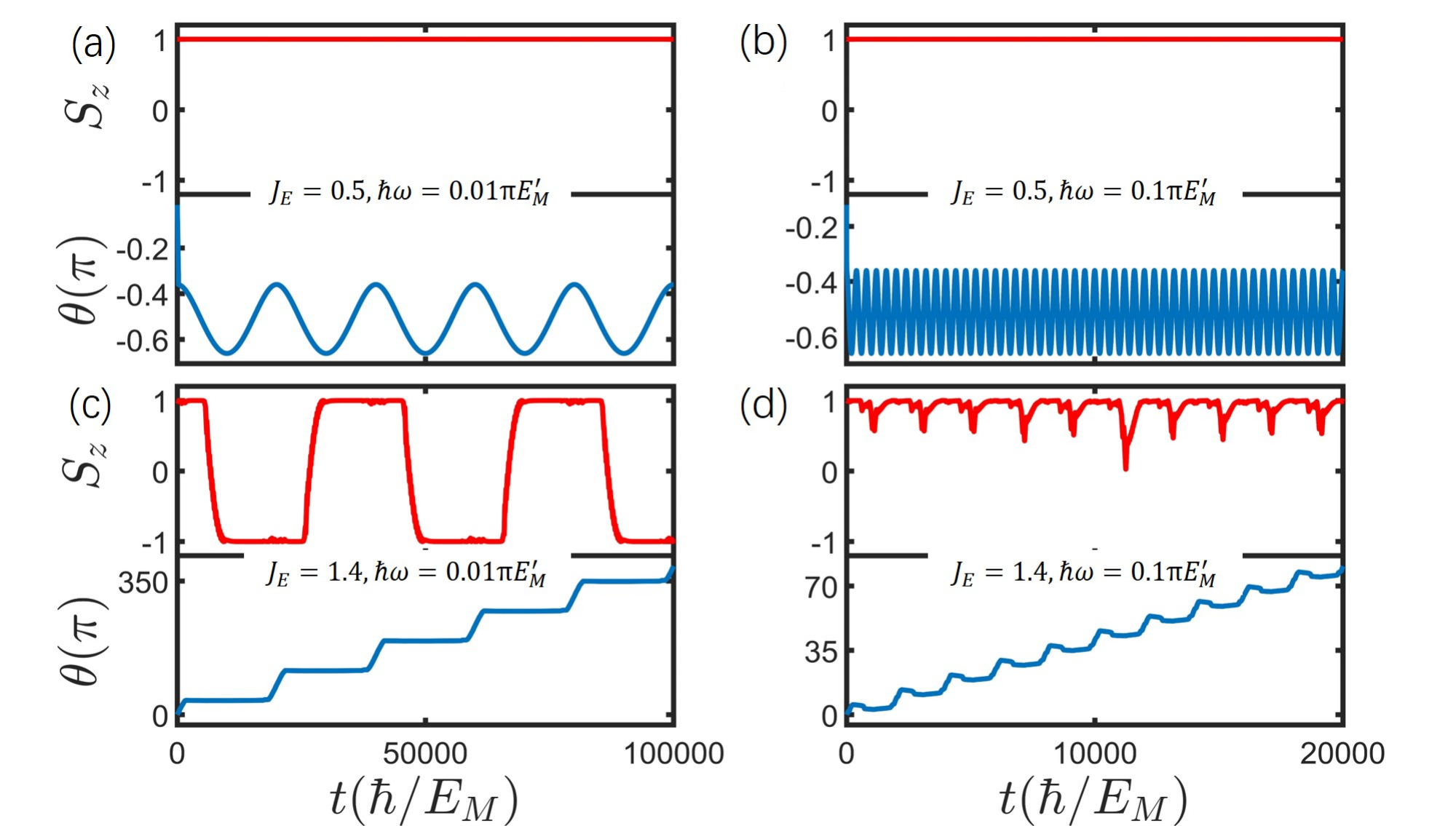}
	\caption{The time evolution of the Josephson phase and the pseudo-spin in the overdamped junction with two typical frequencies and amplitude of the driving ac current. $(a, b)$ The Josephson phase oscillates and average voltage is zero for small driving current. $(c)$ The average voltage is non-zero, while the pseudo-spin evolves adiabatically following the instantaneous ground state at low driving frequency. $(d)$ The average voltage is non-zero, while the pseudo-spin almost stays at the initial state for high driving frequency. Parameters are taken as $I_{c2}/I_{c1}=0.6$, ${16e^{2}I_{c1}}/{E_{J}^{2}C}=1$,${E_{J}}/{2eI_{c1}R}=0.1$, $\varphi/\varphi_{0}={2\pi}/{3}$, $\mathbf{B}_{f}=0.01\hat{\mathbf{x}}$, $\tilde{\mathbf{N}}=\textrm{0.001}\mathbf{I}$, where $\mathbf{I}$ is the identity matrix.}
	\label{fig:3}
\end{figure}

\section{Inverse ac Josephson effect }
\label{section:three}
The equation of motion contains many rich phenomena when the driving force is switched from the dc driving current to the ac driving current. In particular, one could expect an inverse ac Josephson effect where an effective dc voltage is induced by a pure ac driving current. This effect can be understood by examining the I-V curve shown in Fig. \ref{fig:2}a. The critical currents for the opposite directions are asymmetric. Therefore, a near-zero frequency ac driving current with amplitude in between these two critical currents would induce a voltage in one direction while flowing non-disippatively in the other direction. Then a long time averaging over the time evolution of the voltage would give a finite dc component, which is biased in the direction with smaller critical current. In principle, this inverse ac Josephson effect should be a universal effect for Josephson diodes\cite{Margarita2022Science,Liu2024PhysRevB,Daido2022PhysRevLett.128.037001}.

This picture is obtained for near-adiabatic ac driving current. Realistic ac driving currents with reasonable frequencies can be obtained by numerically solving the Eq. (\ref{eq:action_current}) and Eq. (\ref{eq:action_spin}). To be specific,  
we consider an external pure ac current of $I_{\rm ex}=J_E\cos(\omega t)$ in Eq.~(\ref{eq:action_current}) with $J_E$ and $\omega$ are the parameters which control the amplitude and frequency of the input current, and then calculate the time evolution of the Josephson phase $\theta$ and the pseudo-spin $\bf s$ with the standard fourth-order Runge-Kutta method. After a sufficient time evolution, we integrate the Josephson phase over time to obtain the averaged dc voltage as $\bar{V}=\frac{1}{T}\intop_{0}^{T}V(t) dt=\frac{1}{T}\intop_{0}^{T}(\hbar\dot\theta(t)/2e)dt ={\hbar(\theta(T)- \theta(0))}/{2eT}$. With the procedure, we can variable the amplitude and the frequency of the external ac current and obtain the average dc voltage response from the junction.

\begin{figure}
	\centering
	\includegraphics[width=1\columnwidth]{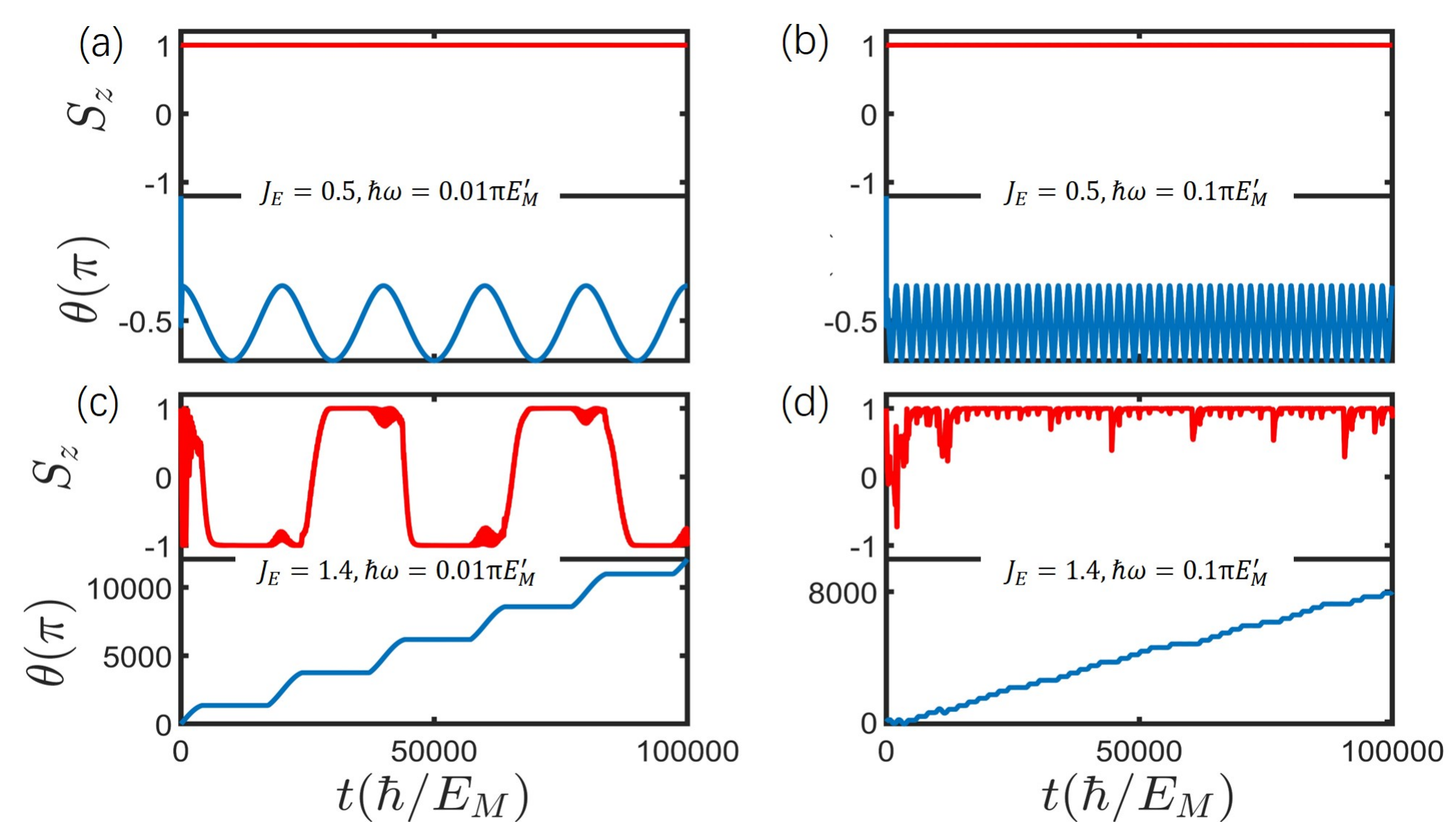}
	\caption{The time evolution of the Josephson phase and the pseudo-spin in the underdamped junction with two typical frequencies and amplitude of the driving ac current. $(a, b)$ The Josephson phase oscillates and average voltage is zero for small driving current. $(c)$ The average voltage is non-zero, while the pseudo-spin evolves adiabatically following the instantaneous ground state at low driving frequency. $(d)$ The average voltage is non-zero, while the pseudo-spin almost stays at the initial state for high driving frequency. The parameters are taken as ${e^{2}I_{c1}}/{E_{J}^{2}C}=4$,${E_{J}}/{2eI_{c1}R}=1$, while other parameters are taken the same as Fig.~\ref{fig:3}.}
	\label{fig:4}
\end{figure}

First, we consider the overdamped junction with a quality factor of $Q=R\sqrt{I_{c1}C}$ smaller than one. In this regime, the dynamics of the Josephson phase can be understood by the motion of mass-less particle whose equation of motion is actually a first-order differential equation. This simplifies the dynamics significantly. We demonstrate the time evolution of the Josephson phase and the pseudo-spin for four typical parameters in Fig. \ref{fig:3}. We first examine the injected current with a low frequency and an amplitude smaller than the maximum current of the conventional junction, that is, $J_E \le I_{c1}$. The injected current is fully transported by the supercurrent channel, and the Josephson phase is purely oscillatory as shown in Fig. \ref{fig:3}a, with zero averaged dc voltage. When the injected current is switched to a higher frequency, the Josephson phase oscillates with a higher frequency as shown in Fig. \ref{fig:3}b. However, the qualitative feature remains unchanged. The averaged dc voltage is zero. 
However, for a larger current, the average dc voltage begins to appear. As shown in Fig. \ref{fig:3}c, a current slightly larger than the maximum current of the conventional channel is able to drive the Josephson phase increasing over time, and the averaged dc voltage becomes non-zero. The injected current has a low frequency; therefore, the evolution of the Majorana qubit is near adiabatic. We notice that the pseudo-spin switches direction, which happens at each avoided crossing for the two eigenenergies of the qubit. This is exactly the adiabatic behavior of the Landau-Zener transition\cite{Feng2020PhysRevB.101.180504,Shevchenko2010}. When the ac frequency of the injected current becomes large, as shown in Fig. \ref{fig:3}d, the pseudo-spin stays in one direction, which means that the probability of the Landau-Zener transition at the avoided crossings is near unity. This is the non-adiabatic limit for the dynamics of the Majorana qubit. Although the Majorana qubit experiences distinct dynamical processes, the dynamical behavior of the Josephson phase is similar. It increases over time, giving a finite average dc voltage. These results suggest that the inverse ac Josephson effect is a universal behavior of the Josephson diode.

For generality, we also explore the underdamped regime where the quality factor is larger than one. In this regime, the Josephson phase has significant capacitance, which brings an effective inertia to the dynamics of the Josephson phase. At first glance, this extra capacitance should complicate the dynamical behaviors of the junction. However,in contrast to this simple picture, a detailed numerical analysis shows that the additional capacitance does not add extra complexity to the dynamics of the Josephson phase. In fact, it simplified the dynamical behavior of the pseudo-spin. We show the time evolution of the Josephson phase and the pseudo-spin of an underdamped junction in Fig. \ref{fig:4}. As the same for the overdamped scenario, we consider four typical parameters for the injected current. When the applied current is less than the maximum current value of the conventional junction, $s_z$ stays at the initial state, with Josephson phase oscillates as shown in Fig. \ref{fig:4}a and Fig. \ref{fig:4}b. The averaged dc voltage is zero, which is the same as the result of the overdamped junction.When the current magnitude increases beyond the maximum current value of the conventional channel, as shown in Fig. \ref{fig:4}c and Fig. \ref{fig:4}d, the Josephson phase begins to increase with time, giving a finite average dc voltage. This also looks similar to the behavior of the overdamped junction. However, the averaged voltages are entirely different, since the Josephson phase increases much faster. 

Now the near-adiabatic behavior at the low driving frequency disappears, $s_z$ no longer oscillates between the opposite directions but stays in one direction. This is due to the effective mass of the Josephson phase coming from the capacitance, which brings an inertia that accelerates the velocity of the Josephson phase so that it will reach the non-adiabatic regime even for a driving current with a small frequency.

\begin{figure}
	\centering
	\includegraphics[width=0.9\columnwidth]{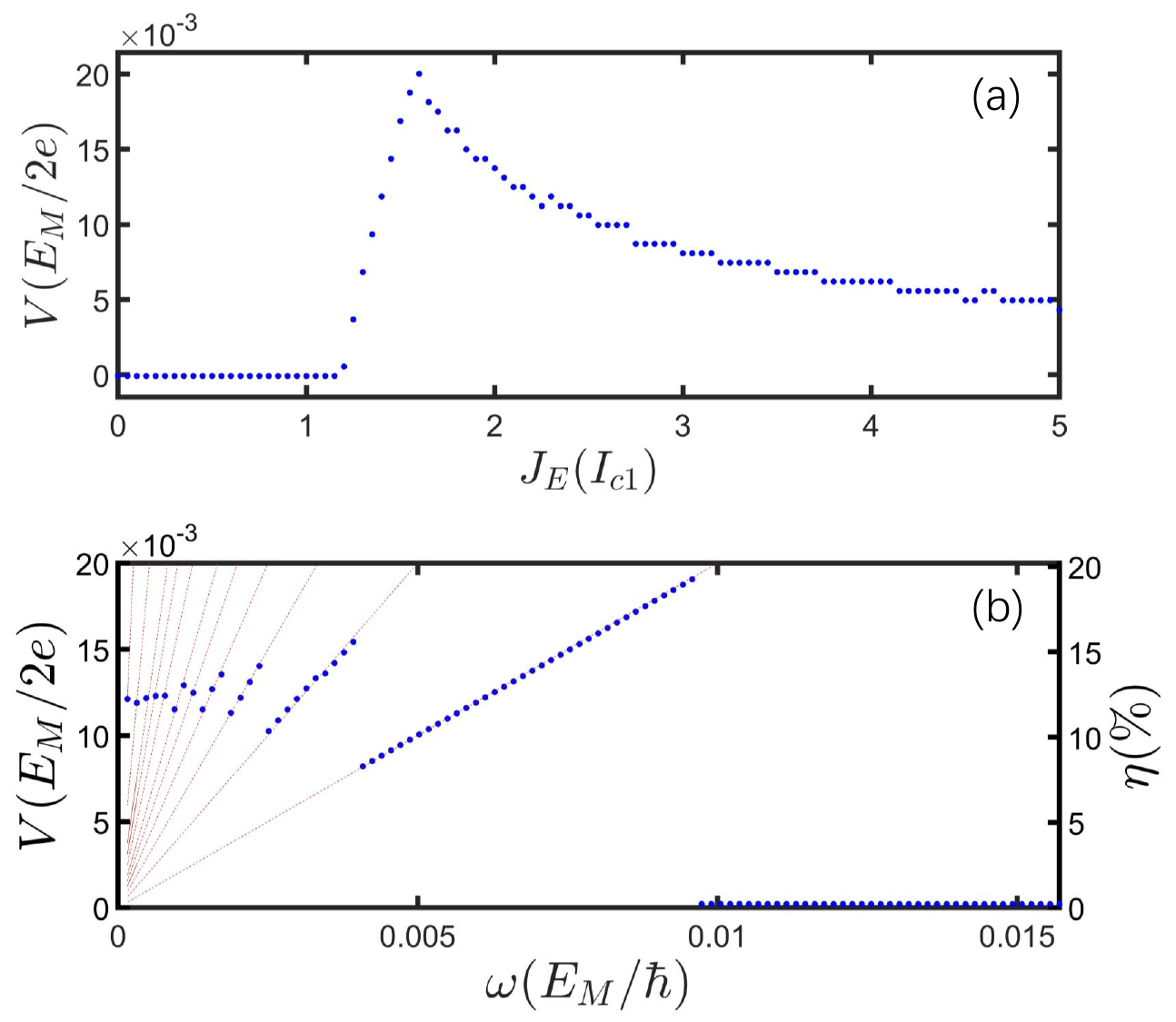}
	\caption{ $(a)$ The average voltage as a function of the magnitude of the applied ac current. Non-zero average voltage appears after a threshold of the driving current value. $(b)$ The average voltage as a function of the frequency of the applied ac current. At low frequency, the behavior mimics the ac Josephson relation. The parameters are taken the same as Fig. \ref{fig:3}.}
	\label{fig:5}
\end{figure}

\section{Driving parameters and efficiency}
\label{section:four}
Now we explore the dependence of the average voltage under different driving currents. We calculate the the average voltage as a function of the amplitude of the ac driving current $I_{ex}$ and show the results in Fig.~\ref{fig:5}a. We find that for ac driving current that is small that the lower critical current, the average voltage is zero. When the driving current exceeds the lower critical current, the 
average voltage suddenly jumps to a finite value, and then gradually decreases with increasing current amplitude. 

Then we consider the average voltage as a function of the frequency of the driving ac current $\omega$, as shown in Fig.~\ref{fig:5}b. We find that the average voltage is zero above certain critical frequency. Below the critical frequency, the average voltage obeys the relation $2eV=n\hbar \omega$ which was marked as red dotted lines. This is actually identical to the ac Josephson relation but with a reversal between the driving force and the response signal. For studying the physical effect about Josephson diode, it would be informative to discuss an effective efficiency. In current study, one of the method to define a efficiency is to take the ratio between the generated dc voltage $V$ and the effective input ac voltage $J_E R/\sqrt{2}$. Then we demonstrate the efficiency of the inverse Josephson as a function of the frequency of the driving ac current $\omega$, as shown in Fig.~\ref{fig:5}b. We find that the efficiency is considerable for small frequencies. To maximize the efficiency one should take proper frequency for the driving current.

\begin{figure}
	\centering
	\includegraphics[width=0.9\columnwidth]{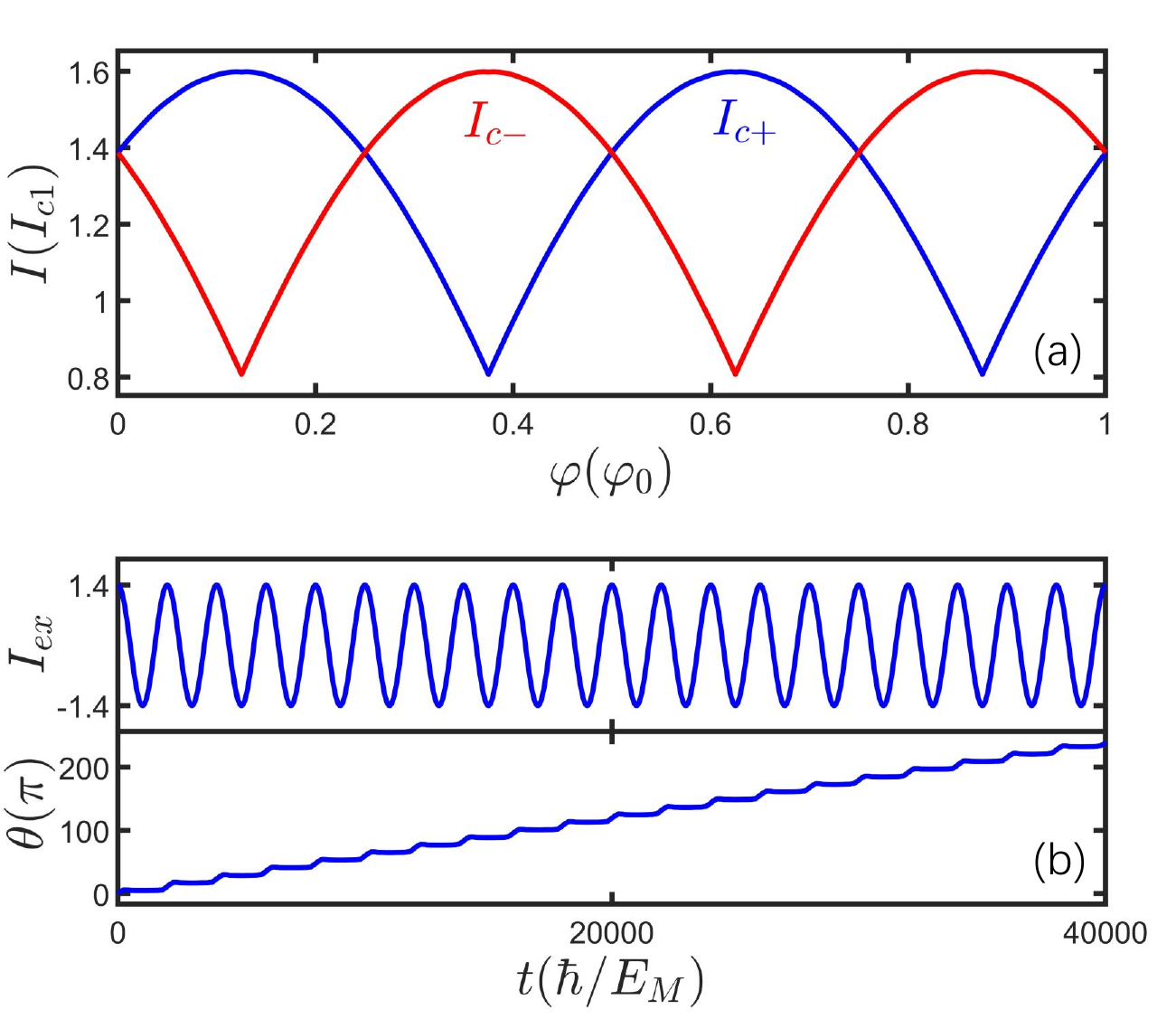}
	\caption{ $(a)$ The absolute value of the critical current for the two opposite directions. $(b)$ The time evolution of the ac driving current and the Josephson phase. The Josephson phase increases with time signaling a dc voltage. The parameters are taken the same as Fig. \ref{fig:3}.}
	\label{fig:6}
\end{figure}

\section{Universality of the effect}
In the previous section we took the one dimensional topological junction as the model for a realistic Josephson diode, however, the Majorana zero modes might be fragile and can only survive in a very narrow energy region. For a driving current with high frequency, the diode effect might be significantly suppressed and the proposed signal might be difficult to observe.

On the other hand, we note that the inverse ac Josephson effect proposed in the this work is actually universal for Josephson diode. Its existence relies on the nonreciprocal critical current instead of the Majorana zero mode. To demonstrate this we consider an even simpler model which is mathematical but intuitive. One of simplest model for Josephson diode is the combination of Josephson currents with different harmonics. We take the junction with a second harmonic as an example, the current phase relation is written as $I=I_{c1}\sin(\theta+\varphi)+I_{c2}\sin(2{\theta})$. This current phase relation exhibits nonreciprocal critical currents shown as Fig. \ref{fig:6}a. For this junction, the dynamics of the Josephson phase is determined by the textbook resistively shunted junction model,
\begin{equation}
 I_{ex} = \frac{\dot{\theta}}{R}+I_{c1}\sin(\theta+\varphi)+I_{c2}\sin(2{\theta}).
\end{equation}
For this system we implement a pure ac driving current. As shown in Fig. \ref{fig:6}b, the Josephson phase increases with time and a dc voltage is generated. This result demonstrate the universality of the proposed inverse ac Josephson effect.

\section{Conclusion}
In conclusion, we study the Josephson diode constructed by the conventional junction and a topological junction. With the quantum resistively and capacitively shunted junction model, we show that the nonreciprocal critical current of the junction induces an inverse ac Josephson effect, where a pure ac driving current leads to a average dc voltage. We study the dynamics of the Josephson phase and demonstrate the dependence of the average voltage upon the driving ac current.

\section*{Acknowledgments} This work was supported
by NSFC (Grants No.12174453 and No.92165204), and the Guangdong Provincial Key Laboratory of Magnetoelectric Physics and Devices No. 2022B1212010008.

\bibliographystyle{apsrev4-2}
\bibliography{ref.bib}

\end{document}